\documentclass[preprint]{elsarticle}
\usepackage[margin=1.25in]{geometry}

\usepackage{tikz}

\usepackage[utf8]{inputenc} 
\usepackage[T1]{fontenc}    
\usepackage{hyperref}       
\usepackage{url}            
\usepackage{booktabs}       
\usepackage{amsfonts}       
\usepackage{nicefrac}       
\usepackage{microtype}      
\usepackage{lipsum}
\usepackage{listings}
\usepackage{graphicx}
\usepackage{color}
\usepackage{soul}
\usepackage{mathtools}
\usepackage{comment}
\usepackage{caption}

\usetikzlibrary{positioning}

\usepackage{bigfoot}

\usepackage{amsmath}
\usepackage{bm}
\usepackage{float}

\usepackage{multirow}
\newcommand{\mr}[2]{\multirow{#1}{*}{#2}}
\newcommand{\mc}[3]{\multicolumn{#1}{#2}{#3}}

\journal{Computer Physics Communications}

\begin{document}
\title{
Massive-Scale Simulations of 2D Ising and Blume-Capel Models on Rack-Scale Multi-GPU Systems
}

\author[1]{Mauro Bisson}
\author[2]{Massimo Bernaschi}
\author[1]{Massimiliano Fatica}
\author[3]{Nikolaos G. Fytas}
\author[2]{Isidoro Gonz\'{a}lez-Adalid Pemart\'{i}n}
\author[4]{V\'{i}ctor Mart\'{i}n-Mayor}
\author[3]{Alexandros Vasilopoulos}
\address[1]{NVIDIA Corporation Santa Clara, CA 95051, USA}
\address[2]{Istituto per le Applicazioni del Calcolo, CNR - Via dei Taurini 19, 00185 Rome, Italy}
\address[3]{School of Mathematics, Statistics and Actuarial Science, University of Essex, Colchester CO4 3SQ, United Kingdom}
\address[4]{Departamento de F\'isica T\'eorica, Universidad Complutense, 28040 Madrid, Spain}

\def\thefootnote{*}

\begin{abstract}

We present high-performance implementations of the two-dimensional Ising and Blume-Capel models for large-scale, multi-GPU simulations. Our approach takes full advantage of the NVIDIA GB200 NVL72 system, which features up to $72$ GPUs interconnected via high-bandwidth NVLink, enabling direct GPU-to-GPU memory access across multiple nodes. By utilizing Fabric Memory and an optimized Monte Carlo kernel for the Ising model, our implementation supports simulations of systems with linear sizes up to $L=2^{23}$, corresponding to approximately $70$ trillion spins. This allows for a peak processing rate of nearly $1.15 \times 10^5$ lattice updates per nanosecond—setting a new performance benchmark for Ising model simulations. Additionally, we introduce a custom protocol for computing correlation functions, which strikes an optimal balance between computational efficiency and statistical accuracy. This protocol enables large-scale simulations without incurring prohibitive runtime costs. Benchmark results show near-perfect strong and weak scaling up to $64$ GPUs, demonstrating the effectiveness of our approach for large-scale statistical physics simulations.

\end{abstract}

\maketitle

\section*{Program Description}

\begin{description}

\item[Program title:] cuIsing (optimized)

\item[Licensing provisions:] MIT license

\item[Programming languages:] CUDA C

\item[Nature of problem:] Comparative studies of the critical dynamics of the Ising and Blume-Capel models are essential for gaining deeper insights into phase transitions, enhancing computational methods, and developing more accurate models for complex physical systems. To minimize finite-size effects and optimize the statistical quality of simulations, large-scale simulations over extended time scales are necessary. To support this, we provide two high-performance codes capable of running simulations with up to 70 trillion spins.

\item[Solution method:] We present updated versions of our multi-GPU code for Monte Carlo simulations, implementing both the Ising and Blume-Capel models. These codes take full advantage of multi-node NVLink systems, such as the NVIDIA GB200 NVL72, enabling scaling across GPUs connected across different nodes within the same NVLink domain. Communication between GPUs is handled seamlessly via Fabric Memory--a novel memory allocation technique that facilitates direct memory access between GPUs within the same domain, eliminating the need for explicit data transfers. By employing highly optimized CUDA kernels for the Metropolis algorithm and a custom protocol that reduces the computational overhead of the correlation function, our implementation achieves the highest recorded performance to date.

\end{description}


\section{Introduction\label{sec:intro}}

We present computer programs specifically developed to simulate the out-of-equilibrium dynamics of the two-dimensional (2D) Ising and Blume-Capel models—two paradigmatic systems that belong to the same \emph{equilibrium} universality class~\cite{zinn-justin:05,parisi:88,amit:05}. Our chosen dynamics is the standard Metropolis spin-flip algorithm, see, e.g.~\cite{landau:05,sokal:97}, representative of the so-called \emph{model-A} dynamic universality class,  which describes local dynamics without conservation laws~\cite{hohenberg:77}. Out-of-equilibrium dynamics is a central theme in Statistical Mechanics and plays a crucial role in the theoretical, numerical, and experimental study of disordered magnetic systems (see, e.g.,~\cite{vincent:97,young:98,charbonneau:23,dahlberg:24} for review). Importantly, it is also a vibrant area of research for non-disordered magnetic systems as well~\cite{bray:94}, with increasing interest in recent years
(see, e.g.,~\cite{arenzon:07,gal:20,gonzalez-adalid-pemartin:21,gonzalez-adalid-pemartin:24,teza:25,PRLwannabe}). 
It is important to emphasize that the physics studied here is specific to model-A dynamics. Consequently, simulation algorithms belonging to other dynamic universality classes—such as cluster methods~\cite{swendsen:87,wolff:89} or the Worm algorithm~\cite{prokofev:01}—are not suitable for our purposes, despite their well-known efficiency in driving systems to thermal equilibrium.

In a typical experimental or computational setting, a system is quenched at the initial time $t=0$ from a high temperature down to a working temperature $T$, where it is then allowed to relax. The chosen temperature $T$ typically lies close to, or below, the critical temperature $T_\mathrm{c}$. The relaxation process involves the time-dependent growth of magnetic domains, characterized by a linear size $\xi(t)$. In a typical experiment (see, e.g.,~\cite{dahlberg:24}), the domain size remains much smaller than the overall system size, i.e., $L\gg \xi(t)$.

To obtain results that are effectively indistinguishable from those in the large-$L$ limit, the system size should exceed the correlation length by at least a factor of (say) $50$. As a result, in numerical simulations, the system size effectively acts as a cutoff that limits the maximum correlation length $\xi(t)$ attainable during the simulation (because one needs to ensure that $L$ is at least as large as 50 $\xi(t)$ for all times considered in the simulation). In fact, simulating a single large system, rather than performing multiple runs on smaller systems, offers two significant advantages. First, it enables $\xi(t)$ to evolve over a broader range without being affected by finite-size effects. This is particularly important, as $\xi(t)$ serves as the fundamental scale variable in the scaling analyses central to the study of out-of-equilibrium dynamics (see, e.g.,~\cite{PRLwannabe,fernandez:15,janus:18}). Second, statistical errors decrease with increasing system size, scaling as $[\xi(t)/L]^{D/2}$, where $D$ is the spatial dimensionality (with $D=2$ in our case). In practical terms, this means that a single simulation of a system with size 
$L'=2L$ provides statistical precision equivalent to 
$2^D$ independent simulations of a system of size $L$.

The simplicity of spin models makes them particularly well-suited for implementation on dedicated hardware. Indeed, several computers have been purpose-built for the simulation of 2D and 3D spin systems. In parallel, highly specialized software has been developed for platforms such as Graphics Processing Units (GPUs). A portion of this work focuses on simulation algorithms that do not belong to the model-A dynamic universality class—for example, implementations on dedicated machines~\cite{talapov:92,talapov:93,blote:99} and GPU-based approaches~\cite{yavorskii:12,weigel:11b,weigel:12a,weigel:12b,borovsky:16,barash:17}. However, considerable effort has also been invested in the implementation of model-A dynamics, both in hardware platforms~\cite{hoogland:83,pearson:83,andreichenko:91,condon:85,ogielski:85a,talapov:90,cruz:01,janus:08,janus:14} and in GPU-based software solutions~\cite{weigel:11a}.

Most directly related to our work are studies that apply model-A algorithms to simulations of the 2D ferromagnetic Ising model. This model has been simulated on various hardware platforms, including special-purpose machines with Field Programmable Gate Arrays (FPGAs)~\cite{FPGA}, Tensor Processing Units (TPUs)~\cite{TPU2019}, the Cerebras Wafer-Scale Engine (WSE)~\cite{Cerebras}, and Graphics Processing Units (GPUs)~\cite{Ising2020, bernaschi24, Bernaschi2012, Tobias1, Tobias2,block:10}. In this work, we update the highly optimized CUDA implementation of the 2D Ising model previously released in Ref.~\cite{Ising2020}. The two key innovations in this version are: $i)$ the ability to simulate much larger systems than in the previous version (the importance of this point has already been emphasized above), and $ii)$ support for the Blume-Capel model in addition to the standard Ising model. 

Our new implementation of the Ising and Blume-Capel models for multi-GPU clusters is capable of scaling from a single GPU to the latest NVIDIA GB200 NVL72 rack-scale architecture.
It is designed to leverage the direct GPU-to-GPU
communication across and within the nodes to realize a single coherent memory space that can be shared by multiple devices to distribute the computations effectively. Moreover, all communications resulting from devices accessing portions of the memory space residing on remote GPUs are handled transparently through NVLink eliminating the need for explicit memory transfers. 

The rest of this paper is organized as follows: in Sec.~\ref{sec:models}, we review the basic features of the Ising and Blume-Capel models. Section~\ref{sec:implementation} details the new implementation, while Sec.~\ref{sec:results} presents performance results. Finally, the paper concludes in Sec.~\ref{sec:conclusions} with a discussion of possible future developments.

\section{Ising and Blume-Capel Models \label{sec:models}}

The 2D spin-$1/2$ Ising model (IM)~\cite{Ising}, often called the \emph{fruit-fly} model of statistical physics, is described by the Hamiltonian~\cite{newman_book}
\begin{equation} \mathcal{H}^{\rm (IM)} = -J\sum_{\langle\mathbf{x},\mathbf{y}\rangle} \sigma_{\mathbf{x}}\sigma_{\mathbf{y}},
\label{eq:IM} 
\end{equation} 
where $\sigma_{\mathbf{x}} = \pm 1$ represents the spins at the nodes of an $L \times L$ square lattice with periodic boundary conditions. The summation $\langle\mathbf{x}, \mathbf{y}\rangle$ runs over nearest neighbors, and $J > 0$ is the ferromagnetic exchange coupling.

In contrast, the spin-$1$ Blume-Capel (BC) model~\cite{blume:66,capel:66}, a generalization of the Ising ferromagnet, offers a promising platform for investigating critical and tricritical phenomena. It has been applied to a broad range of physical systems, from Mott insulators~\cite{lanata:17} to multi-component fluids~\cite{wilding:96}. Its Hamiltonian is given by 
\begin{equation}\label{eq:BC} \mathcal{H}^{\rm (BC)} = -J\sum_{\langle \mathbf{x}, \mathbf{y} \rangle} \sigma_{\mathbf{x}} \sigma_{\mathbf{y}} + \Delta \sum_{\mathbf{x}} \sigma_{\mathbf{x}}^{2}, 
\end{equation} 
where, as in the Ising model, $J > 0$. The key differences in the Blume-Capel model stem from the possibility of vacancies, so that the spin components can take values $\sigma_\mathbf{x} \in \{-1, 0, +1\}$, and from the presence of a chemical potential $\Delta$, known as the crystal-field coupling, which controls the density of vacancies ($\sigma_\mathbf{x} = 0$). For $\Delta \to -\infty$, vacancies are suppressed, and the model reduces to the Ising ferromagnet~\eqref{eq:IM}. The phase boundary of the Blume-Capel model in the crystal-field-temperature ($\Delta$, $T$) plane separates the ferromagnetic and paramagnetic phases~\cite{zierenberg:17}, featuring a tricritical point at $(\Delta_{\rm t}, \; T_{\rm t}) = [1.9660(1),\; 0.6080(1)]$~\cite{kwak:15}. This point distinguishes first-order transitions (at low $T$ and high $\Delta$) from second-order transitions (at high $T$ and low $\Delta$). In particular, second-order transitions follow the Ising universality class~\cite{zierenberg:17, kwak:15,silva:06,malakis:09,malakis:10,fytas:11,zierenberg:15}.

In Ref.~\cite{PRLwannabe}, we compared the critical dynamics of these two models belonging to the same \emph{static} universality class: the square-lattice Ising model and the Blume-Capel model, specifically in its second-order transition regime, with a focus on the critical point corresponding to $\Delta = 0$. In this work, we provide a detailed description of the numerical framework that enabled us to reach those groundbreaking physical conclusions, achieved through simulations effectively at the thermodynamic limit.

\section{Implementation \label{sec:implementation}}

As the foundation for this work, we build upon the \emph{optimized} GPU implementations presented in Ref.~\cite{Ising2020}, which is a single-node/multi-GPU code that employs multi-spin coding~\cite{jacobs:81} (MSC). We also leverage CUDA Unified Memory to enable direct memory access between devices connected to the same machine via high-speed NVLink, eliminating the need for explicit data transfers.

The total size of the systems that can be simulated depends on two factors: the
number of spins that can be stored in the memory of each GPU and the total
number of GPUs available for the simulation. To increase the system size per GPU
in the Ising model, we reduced the bits-per-spin count from $4$ to $1$,
effectively quadrupling the lattice size that can be stored within the same
memory capacity. In contrast, for the Blume-Capel model, we retained the 4-bit
spin representation, as each spin can take three distinct values. We found that
a 2-bit spin representation does not provide an advantage from the performance
viewpoint since the operations remain the same as in the case of the 4-bit
representation.

Additionally, to increase the pool of GPUs available for a single simulation, we modified the code to enable scaling across multiple nodes. This required replacing Managed Memory with Fabric Memory, a new type of allocation that allows memory buffers to span multiple nodes while providing functionality equivalent to Managed Memory on single-node systems.

Reducing the Ising spin representation to a single bit enabled further optimization of the Monte Carlo kernel using a technique described in Ref.~\cite{bernaschi24}, which employs a lookup table to generate non-uniform random bits efficiently. In our original implementation, a uniform random number was used per spin to determine flips based on the acceptance probability. By leveraging the lookup table approach, a single uniform random number can now generate multiple non-uniform random bits, enabling parallel flipping of multiple spins. This strategy significantly reduces the number of random numbers required, addressing one of the most significant bottlenecks in Ising model simulations. However, for the Blume-Capel model, the probability exponential function is considerably more complex than in the Ising case, making the lookup table approach impractical.

Finally, we designed a custom protocol, applicable to both models, for computing the correlation function that reduces computational burden while preserving statistical significance. At specific time steps during the simulation, we compute the spin-spin correlation \emph{from each lattice site} at multiple distances along the two lattice dimensions. Ideally, this would be computed for all distances up to half the system's linear size, $L/2$, but doing so would incur a computational cost of $O(L^3)$, leading to an impractical increase in computation time for large values of $L$. In our protocol, we downsample the correlation function using two distance thresholds. The first sets the maximum distance for which the correlation is computed from each site, while the second specifies the point beyond which sampling follows a logarithmic scale, extending up to $L/2$.

\subsection{Workload Distribution via Fabric Memory}

As is common practice, the spin system is stored in memory using two separate buffers arranged in a checkerboard pattern.\footnote{The checkerboard lattice decomposition is crucial in numerical simulations and pure theoretical settings.  For instance, the lattice decomposition in two sublattices played a significant role in~\cite{barma:77}.} This partitioning ensures that all spins of one color can be updated in parallel. Since each spin's flipping decision depends entirely on its neighbors in the opposite buffer, each simulation step consists of two update phases---one for each color. The workload is distributed across GPUs according to a 1D decomposition, that is by partitioning each buffer into horizontal slabs, which are stored in device memory. This approach minimizes inter-GPU communication during spin updates, restricting it to just two rows: the last row of the slab on the preceding GPU and the first row of the slab on the next GPU. Notably, due to the periodic boundary conditions, the first and last GPU slabs are also connected. In our original code, this slab-based partitioning was easily implemented via Unified Memory. For each color, the host process simply called \verb|cudaMallocManaged| once to request an allocation for the entire buffer. Then, \verb|cudaMemAdvise| was called once per GPU to assign each slab (a specific address range) to its corresponding device's global memory. This allows the GPUs to directly access any spin in the system by indexing the global buffers. The system automatically handles data transfers between devices by migrating memory pages as needed.

While Managed Memory is an effective method for achieving multi-GPU parallelism, it is limited to single-node systems, as it relies on underlying operating system support to provide its functionalities. Fortunately, the features of Unified Memory that our code depends on can be implemented using the Virtual Memory Management API (VMM)\footnote{For a complete description, see: \verb|https://docs.nvidia.com/cuda/cuda-driver-api/group__CUDA__VA.html|.}, a low-level interface for directly managing the GPU's virtual address space. Specifically, VMM introduces a new type of allocation called \emph{fabric memory}, which, in multi-node NVLink systems, allows the memory of remote GPUs to be mapped into the address space of any other GPU.

VMM API calls can be seamlessly integrated into an existing program to configure memory allocations that require low-level tuning, without modifying other parts of the code. For instance, they can be used alongside standard runtime functions like \verb|cudaMalloc| or \verb|cudaMallocManaged|. This flexibility allowed us to introduce support for Fabric Memory in our code by simply replacing the two managed memory calls used for allocating the two color buffers (and the corresponding Unified Memory advise calls) with calls to a custom allocation function. This function encapsulates all the necessary VMM API calls to implement the shared buffer, keeping these details separate from the main code. Moreover, it allows for switching between the single-node/managed-memory and multi-node/Fabric-Memory versions using a compile-time directive that determines which allocation calls to use. Finally, to enable multi-node execution, we used the MPI library to launch and coordinate processes across the nodes. Each MPI process was assigned to a node, where it managed all GPUs connected to that node, following a similar approach to the one used in the original single-node implementation.

The VMM API decouples addresses from memory, enabling applications to manage
them independently and map different memory types and locations to a virtual
address range. Our Fabric Memory function is called by all MPI processes. Each
process first requests a physical memory allocation for each of its local GPUs,
with the allocation size matching the slab size and suitable for sharing with
other GPUs within the same NVLink fabric (both local and remote). This is done
using the \verb|cuMemCreate| function, which returns a handle to the allocated
physical memory. At this point, the memory is reserved but not yet associated
with an address, making it inaccessible. Next, the handles must be distributed
across the processes so that each process can map remote allocations to local
address ranges. However, since handles cannot be shared directly, they are first
exported to OS-specific types using the \verb|cuMemExportToShareableHandle|
function. The exported handles are then distributed using the
\verb|MPI_Allgather| collective. Once received, the remote handles are imported
with \verb|cuMemImportFromShareableHandle|, making them ready for mapping to
local addresses. Each process then reserves a new virtual address range, large
enough to accommodate the entire memory allocation across all devices, using the
\verb|cuMemAddressReserve| function. Next, each memory handle is mapped to
consecutive chunks (each of slab size) within the newly reserved address range
using \verb|cuMemMap|. Access to the address range, as well as the underlying memory,
is enabled via \verb|cuMemSetAccess|. To ensure that the same offset in the
address range of each process corresponds to the same physical memory location
(on the same GPU), all processes map the handles in a consistent order: first
across MPI ranks (from the first to the last), and then, within each rank, from
the first GPU to the last.

Note that, unlike managed memory, the beginning of the virtual address range \verb|[startPtr,| \verb|startPtr+totSize)| may differ for each process, even though the underlying memory it maps to remains identical. This behavior is common in many shared memory mechanisms where there is a decoupling between memory allocation and addressing. The allocation function then returns \verb|startPtr| to the caller.

\begin{figure*}[t]
\begin{center}
        \includegraphics[scale=1.2]{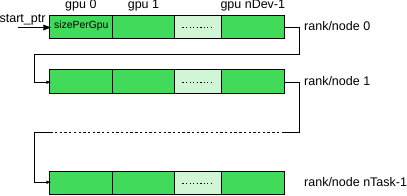}
\end{center}
        \caption{{Distribution of the shared spin buffer across
        the GPUs connected to the nodes}.}
        \label{fig:shared_buffer}
\end{figure*}

In this way, we have implemented a shared buffer spanning multiple GPUs across different nodes, as illustrated in Fig.~\ref{fig:shared_buffer}. On each node, the address \verb|startPtr| points to the same contiguous buffer, with consecutive chunks of equal size (matching the size of a spin slab) physically distributed across the memory of consecutive GPUs across all nodes. As a result, \verb|startPtr+i| consistently refers to the same memory location within the memory of a particular GPU, regardless of the process accessing it within the system.

From the application's perspective, there is no difference compared to memory allocation using \verb|cudaMallocManaged|. Each process provides the same buffer pointer to the kernels launched on its local devices, enabling GPU threads to access any location directly. Accesses to locations mapped to remote memory are automatically handled via NVLink transfers, transparently managed by the system. This approach eliminates the need for explicit data transfers between GPUs and maximizes the size of the systems that can be simulated, as remote data can be accessed directly without the need to reserve additional memory for local copies.

\subsection{Optimized Monte Carlo Implementation}

The original implementation of our spin update kernel employed Multiple Spin Coding (MSC) by using four bits per spin to sum the values of neighboring spins for multiple consecutive spins (of the same color) in parallel, within the same addition instructions. In the updated version, we reduced the bits per spin to one in order to maximize the size of the spin slab per device. This required a significant rewrite of the kernel, as integer additions could no longer be used to sum spin values in parallel—any sum involving more than one value of $1$ would overflow into the next site.
As mentioned earlier, using one bit per spin also allowed us to improve kernel efficiency by reducing the number of random numbers required to flip spins according to the acceptance probability. This improvement is achieved through the LUT-based approach described in Ref.~\cite{bernaschi24}, Section II.A.1.
We recall that a spin is flipped with probability $e^{-4\beta}$
if it shares its value with exactly three neighbors, with probability $e^{-8\beta}$ if it shares its value with exactly four neighbors, and with probability $1$ in all other cases.  
Since we are using one bit per spin, we pack multiple spins into a single unsigned word (the word bit-length will be discussed later). 

To explain how we revised the update kernel, let us assume we have the words  \verb|src|, \verb|eq4| and
\verb|eq3|, where (for clarity, we use array indexing to denote a specific bit within a word):
\begin{itemize}
    \item \verb|src| contains the input spins;
    \item \verb|eq4[i] == 1| $\iff$ exactly four neighbors of spin $i$
          have the same value as \verb|src[i]|;
    \item \verb|eq3[i] == 1| $\iff$ exactly three neighbors of spin $i$
          have the same value as \verb|src[i]|.
\end{itemize}
\noindent
Clearly, "\verb|eq4 & eq3|" is equal always to zero. Let us also define a function \verb|nonUniformRandBits()| that returns a word in which each bit is set independently with probability $e^{-4\beta}$. With these definitions, we can
update the source spins in parallel as follows:
\begin{verbatim}
exp4 = nonUniformRandBits();
exp8 = nonUniformRandBits();
exp8 = exp8 & exp4;
flip = (eq4 & exp8) | (eq3 & exp4) | (~eq4 & ~eq3);
dst = src ^ flip;
\end{verbatim}
The words \verb|eq4| and \verb|eq3| can be computed starting from the source spin word and the words containing the four neighboring spins in several ways. Let the neighboring words be labeled \verb|neighN|, \verb|neighS|, \verb|neighW| and \verb|neighE|. We obtain them using the following method:
\begin{verbatim}
neighN ^= ~src;
neighS ^= ~src;
neighW ^= ~src;
neighE ^= ~src;

tmp = neighW; neighW &= neighE; neighE |= tmp;
tmp = neighS; neighS &= neighW; neighW |= tmp;
tmp = neighN; neighN &= neighS; neighS |= tmp;
tmp = neighW; neighW &= neighE; neighE |= tmp;
tmp = neighS; neighS &= neighW; neighW |= tmp;

eq4 =  neighN;
eq3 = ~neighN & neighS;
\end{verbatim}

The first four lines replace the values of the neighboring spins with a bit indicating whether each neighbor matches the corresponding source spin (a value of 1 means they match). Then, we independently sort these ``difference'' bits within the word sequence \verb|[neighN,| \verb|neighS,| \verb|neighW,|
\verb|neighE]|. Specifically, bits with a value of $0$ are compacted toward \verb|neighN| (while bits with a value of $1$ are shifted toward \verb|neighE|). This is accomplished by swapping adjacent bits, starting from the end of the sequence, until the lowest values are moved into the first two elements. Since we are sorting four elements and begin from the end, only five swaps are needed. If a spin shares its value with four neighbors, the corresponding bit in \verb|neighN| is set to $1$. If it shares its value with exactly three neighbors, the corresponding bits in \verb|neighN| and \verb|neighS| are set to $0$ and $1$, respectively. The final two lines store this information in the words \verb|eq4| and \verb|eq3|.

In our code, each GPU thread manages $128$ spins, enabling efficient memory access by performing $128$-bit loads and stores using the \verb|ulonglong2| vector type. Accordingly, we use two $64$-bit unsigned long long integers for each spin word in the above description (specifically, the \verb|.x| and \verb|.y| fields of the vector type).

Finally, the \verb|nonUniformRandBits()| function generates non-uniform random bits. It is based on an algorithm that produces 
$N$ random bits by performing a binary search on a $32$-bit random number within a sorted table of length $2^N$.
The table is generated at the beginning of the run and remains unchanged throughout the simulation. The probability distribution of the bits ($p(0)=1-e^{-4\beta}$ and $p(1)=e^{-4\beta}$)
is encoded in the content of the table. A detailed description of how it is generated can be found in Ref.~\cite{bernaschi24}. In our case, $N=4$, so we generate four random bits per lookup. The function \verb|nonUniformRandBits()| is implemented as follows:
\begin{verbatim}
function nonUniformRandBits() {
    uint64 ret = 0
    for(int k = 0; k < 64; k += 4) {
            uint32 rndNum  = philox_32_10();
            uint64 rndBits = getMaxLE(table, rndNum)
            ret |= rndBits << k;
    }
    return ret
}
\end{verbatim}

The $32$-bit uniform random number is generated using the \verb|Philox4_32_10| generator from the cuRAND library’s device API~\cite{cuRAND}, as in our original code. The function \verb|getMaxLE()| returns the index (from $0$ to $15$) of the table corresponding to the largest element less than or equal to \verb|rndNum|. For each $64$-bit word (representing $64$ spins), we generate $16$ random numbers. Since two random bit words (\verb|exp8| and \verb|exp4|) are needed for each spin word, the total number of calls to the random number generator is $32$. This reduces the number of calls by a factor of $2$ compared to our original implementation, where each spin required a separate $32$-bit random number.

\subsection{Correlation Protocol\label{sec:CorrProt}}

As discussed in Sec.~\ref{sec:intro}, we developed a custom protocol to reduce the computational time for calculating the correlation function without compromising statistical reliability. Ideally, the correlation function would be computed at every site and for all distances $r$ up to half the system's linear dimension. However, this would result in prohibitively long runtimes for large systems. Using a Fast Fourier Transform (FFT) is not a feasible alternative for two reasons. First, in the Blume-Capel model, the FFT would require unpacking the spin configuration. Second, and more critically, we are already close to exhausting the device’s memory with the spin configurations.

Our approach computes the spin-spin correlation for each spin in the system only up to a limited threshold, $r\le 2R$. Beyond this threshold, we apply downsampling. Specifically, we reduce the number of source spins by a factor of $R^2$ by selecting a single spin per $R\times R$ square (e.g., the spin in the top-left corner). This downsampling comes with almost  no penalty in statistical errors as soon as the coherence length $\xi(t)$ exceeds $R$.\footnote{In fact, the number of \emph{statistically independent} source spins in a lattice of dimension $L$ is roughly $[L/\xi(t)]^2$.} Despite this downsampling, calculating the correlation for every distance up to $L/2$ would still result in excessive runtimes for large systems (we simulated systems with a linear dimensions on the order of $O(10^6)$).

To further optimize performance, we introduce an additional level of downsampling by defining a second distance threshold, $r_{\rm c}(t)$, beyond which correlation values are sampled using a logarithmic scale. However, in order to compute the coherence length safely, both $r_c(t)$ and the statistical errors must meet two separate conditions. First, $r_{\rm c}(t)$ must be larger than a fixed multiple of the coherence length $\xi(t)$. Since the growth of $\xi(t)$ is slightly slower than $\sqrt{t}$, we found it sensible to have $r_{\rm c}(t)$ grow linearly with $\sqrt{t}$, as explained below. Additionally, we require full details of the correlation function up to the largest distance where the signal-to-noise ratio remains above a prescribed threshold (3 in our case). With these conditions in mind, statistical errors decay with system size as $1/L$, which implies that $r_{\rm c}(t)$ should increase with system size (though not by much, since the correlation function decays at long distances as 
$C(r)\sim \mathrm{e}^{-[r/\xi(t)]^2}$). Our prescription, described below, was calibrated using preliminary simulations of $100$ independent runs of the Blume-Capel system with sizes 
$L = 2^{16}$. This calibration extends to larger systems, as discussed. Specifically, we compute the correlation function $C(r, t)$ as follows:

\begin{itemize}
\item for all spins, for $r\le 2R$;
\item for one spin per $R\times R$ square, for $2R < r \le r_{\rm c}(t)$;
\item for one spin per $R\times R$ square, for $r \in \{\lfloor 2^{x/32} \rfloor : r_{\rm c}(t) < x \le 32[log_2(L)-1]\}$, meaning $32$ values of $r$ for each power of two, up to $L/2$;
\end{itemize}

\noindent
with:
$$ R= 16, \quad r_{\rm c}(t) = \max\{256, \lfloor g(L) \sqrt{t} + 0.5\rfloor\}, \quad  g(L) = 6 \sqrt{ \frac{\log(L / 2^{16})}{3.3^2} + 1.0 } $$

Even with this protocol, calculating the correlation function at every timestep would still significantly impact the overall simulation runtime. To mitigate this, we compute the function at timesteps spaced evenly on a logarithmic scale, a natural choice for studying a power-law time-growth phenomenon:
$$\{t : t = \lfloor (2^{0.125})^x+0.5\rfloor\,,\quad x\ \mathrm{\ a\ positive\ integer\, }\}.$$
This approach evaluates the correlation function less frequently as the simulation progresses, striking a balance between accuracy and computational efficiency. For instance, in our large Ising simulations presented in Ref.~\cite{PRLwannabe}, we ran $524,288$ timesteps on systems with $L=2^{22}$, with each run lasting a total of $24.1$ hours. The correlation function computation accounted for approximately $3.3\%$ of the total runtime.

\section{Results \label{sec:results}}

In this section, we present performance results for both single-GPU and multi-GPU configurations. Our tests were conducted on an NVIDIA GB200 NVL72 system, which consists of $18$ computing nodes. Each node is equipped with two Grace CPUs (each featuring $72$ ARM Neoverse V2 cores) and four Blackwell GPUs, for a total of $72$ GPUs. All GPUs are connected within a single NVLink domain, providing a GPU-to-GPU bandwidth of $1.8$ TB/s and an aggregate NVLink bandwidth of $130$ TB/s. Each GB200 GPU is equipped with $192$ GB of HBM3e memory, delivering a bandwidth of $8$ TB/s. This results in a total device memory of $13.5$ TB and an aggregated memory bandwidth of $576$ TB/s.

While this work emphasizes the performance of our Ising and Blume-Capel implementations, these codes have already been applied in large-scale numerical studies demonstrating the universality of critical dynamics. The results of these studies, available in Ref.~\cite{PRLwannabe}, confirm the correctness of our implementations. Additionally, we have used the Schwinger-Dyson equation as a practical and straightforward program sanity check~\cite{ballesteros:98c}. For both the Ising and Blume-Capel models, the following identity holds in thermal equilibrium:
\begin{equation}\label{eq:SDI}
1= \frac{1}{L^2}\sum_{\mathbf{x}}\langle \mathrm{e}^{-2J\beta \sigma_{\mathbf{x}} h_{\mathbf{x}}}\rangle\,,\quad \beta=\frac{1}{k_{\mathrm{B}}T}\,,\quad h_{\mathbf{x}}=\sum_{\Vert \mathbf{x}-\mathbf{y}\Vert=1}\,\sigma_{\mathbf{y}}\,.
\end{equation}
The identity was originally derived for the Ising model, but it straightforwardly extends to the Blume-Capel model. Note that evaluating the right-hand side (r.h.s.) of the above equation is similar to an energy computation using a short look-up table. Specifically, the exponential term takes only five distinct values for the Ising model and nine for the Blume-Capel model. Although our simulations do not run long enough to reach full thermal equilibrium, Eq.~\eqref{eq:SDI} holds with very high accuracy as soon as $\xi(t)$ becomes significantly larger than one (i.e., when the system reaches quasi-equilibrium locally).\footnote{After just $128$ time steps, we observed deviations from Eq.~\eqref{eq:SDI} on the order of $\sim 5\times 10^{-4}$ for $L=2^{23}$ Ising systems, and as small as $2\times 10^{-5}$ for $L=2^{23}$ Blume Capel systems (at their respective critical temperatures).} Failure to meet this condition is a clear indication that something is wrong with the simulation.

We assess the performance improvement of our current Monte Carlo implementation by comparing it to the previous version released in~\cite{Ising2020}. Both codes were executed on a single GB200 GPU, using the largest system size supported by the earlier implementation, $L=2^{19}$. 
That version stored $4$ bits per spin, requiring $128$ GB of memory for the full lattice. The previous code achieved a speed of $1070$ updates/ns, while the new implementation reaches 1800 updates/ns --- an improvement of approximately $1.7\times$.

Tables~\ref{tbl:single_gpu} and \ref{tbl:single_gpu_BC} present the single-GPU performance of the Ising and Blume-Capel codes on an NVIDIA GB200 GPU, varying the system's linear size and covering total memory usage from $8$ MB to $128$ GB. The data from the tables are plotted in Figures \ref{fig:single_gpu} and \ref{fig:single_gpu_BC}. For the Ising code, experiments were conducted with system sizes ranging from $L=2^{13}$ to $L=2^{20}$. Since each spin in the Blume-Capel code requires four bits instead of one, the runs for this case started at $L=2^{12}$ and went up to $L=2^{19}$. Peak performance for the Ising code is reached at $L=2^{16}$ ($1.0$GB), exceeding $99\%$ of the absolute maximum performance of $1802$ updates/ns, which is achieved at $L=2^{20}$. For the Blume-Capel code, peak performance occurs at $L=2^{16}$, exceeding $99\%$ of the maximum performance of $948$ updates/ns, which is attained at $L=2^{19}$.

Table~\ref{tbl:strong_scaling_both} summarizes the strong scaling measurements for both codes. For each model, we ran the same system using between $1$ and $64$ GPUs over $128$ time steps. The lattice sizes were selected to maximize memory usage on a single GPU, with $2^{38}$ spins ($L=2^{19}$) for the Blume-Capel model and $2^{40}$ spins ($L=2^{20}$) for the Ising model. Each experiment produced identical numerical results (within each model), regardless of the number of GPUs used. Figure \ref{fig:strong_scaling_both} illustrates the corresponding speedup graph. The codes demonstrate near-perfect linear speedup, benefiting from the minimal inter-GPU communication required during the spin update step (each GPU exchanges $0.25$ MB and $0.5$ MB of data with its two neighbors for the Ising and Blume-Capel models, respectively).

\begin{table}[h]
    \centering
    \begin{minipage}{\linewidth}
        \centering
        \begin{tabular}{ c c }
            \toprule
                  $L$ & updates/ns  \\ \cmidrule{1-2}
            $2^{13}$  &       1051  \\
            $2^{14}$  &       1530  \\
            $2^{15}$  &       1741  \\
            $2^{16}$  &       1784  \\
            $2^{17}$  &       1797  \\
            $2^{18}$  &       1801  \\
            $2^{19}$  &       1802  \\
            $2^{20}$  &       1802  \\
            \bottomrule
        \end{tabular}
        \caption{Spin updates per nanosecond for the Ising code on a single
        GB200 GPU, with varying system sizes and memory requirements ranging
        from $8$ MB to $128$ GB.}
        \label{tbl:single_gpu}
    \end{minipage}
    \begin{minipage}{0.75\linewidth}
        \centering
         \includegraphics[width=\linewidth]{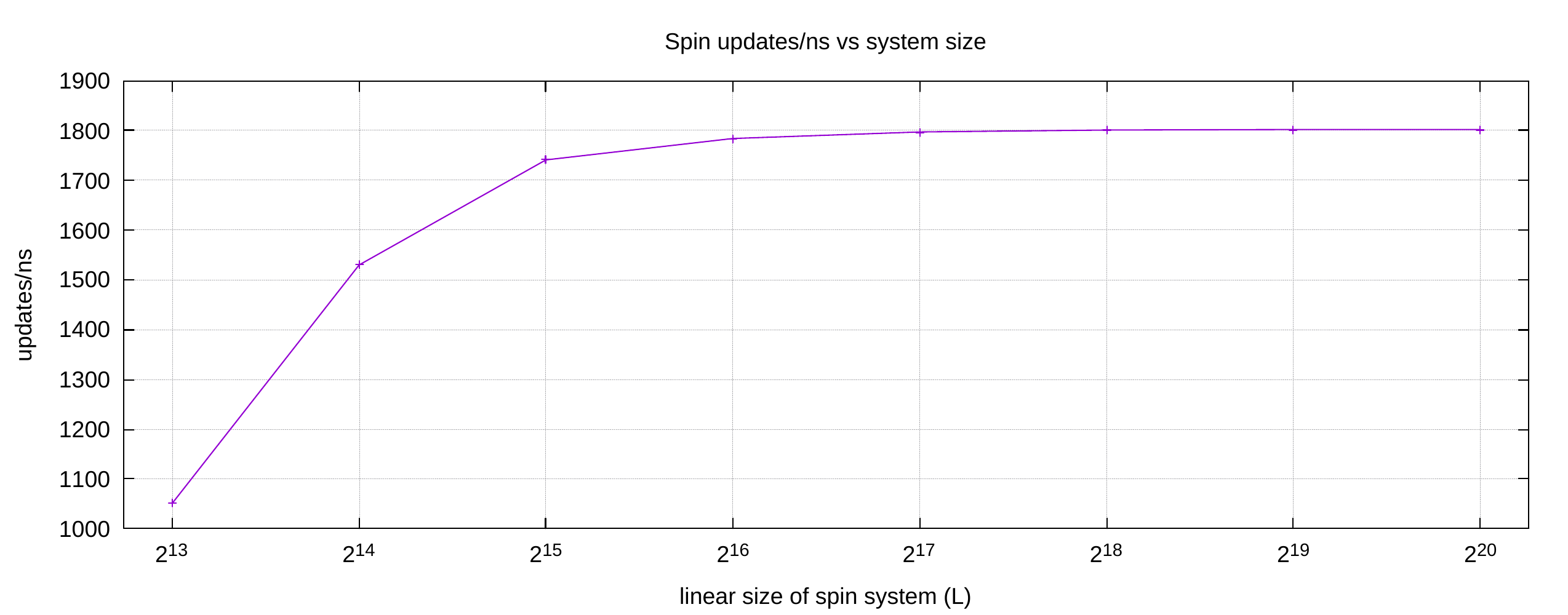}
    \end{minipage}
    \captionof{figure}{Plot showing the spin update throughput data from Table~\ref{tbl:single_gpu}.}
    \label{fig:single_gpu}
\end{table}

\begin{table}[h]
    \centering
    \begin{minipage}{\linewidth}
        \centering
        \begin{tabular}{ c c }
            \toprule
                  $L$ & updates/ns  \\ \cmidrule{1-2}
            $2^{12}$  &        304  \\
            $2^{13}$  &        633  \\
            $2^{14}$  &        839  \\
            $2^{15}$  &        922  \\
            $2^{16}$  &        941  \\
            $2^{17}$  &        947  \\
            $2^{18}$  &        949  \\
            $2^{19}$  &        948  \\
            \bottomrule
        \end{tabular}
        \caption{Spin updates per nanosecond for the Blume-Capel code on a single
        GB200 GPU, with varying system sizes and memory requirements ranging from
        $8$ MB to $128$ GB.}
        \label{tbl:single_gpu_BC}
    \end{minipage}
    \begin{minipage}{0.79\linewidth}
        \centering
        \includegraphics[width=\linewidth]{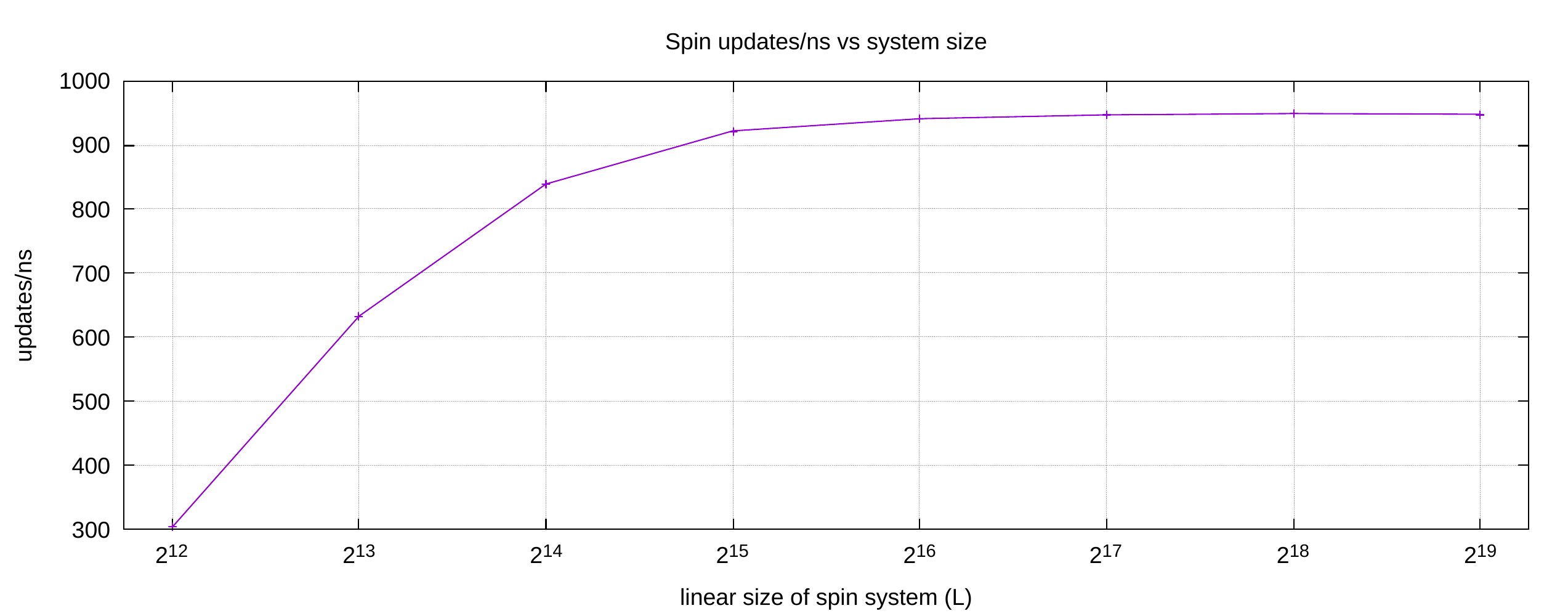}
    \end{minipage}
    \captionof{figure}{Plot showing the spin update throughput data from Table~\ref{tbl:single_gpu_BC}.}
    \label{fig:single_gpu_BC}
\end{table}

\begin{table}[h]
    \centering
    \begin{minipage}{\linewidth}
        \centering
        \includegraphics[scale=0.4]{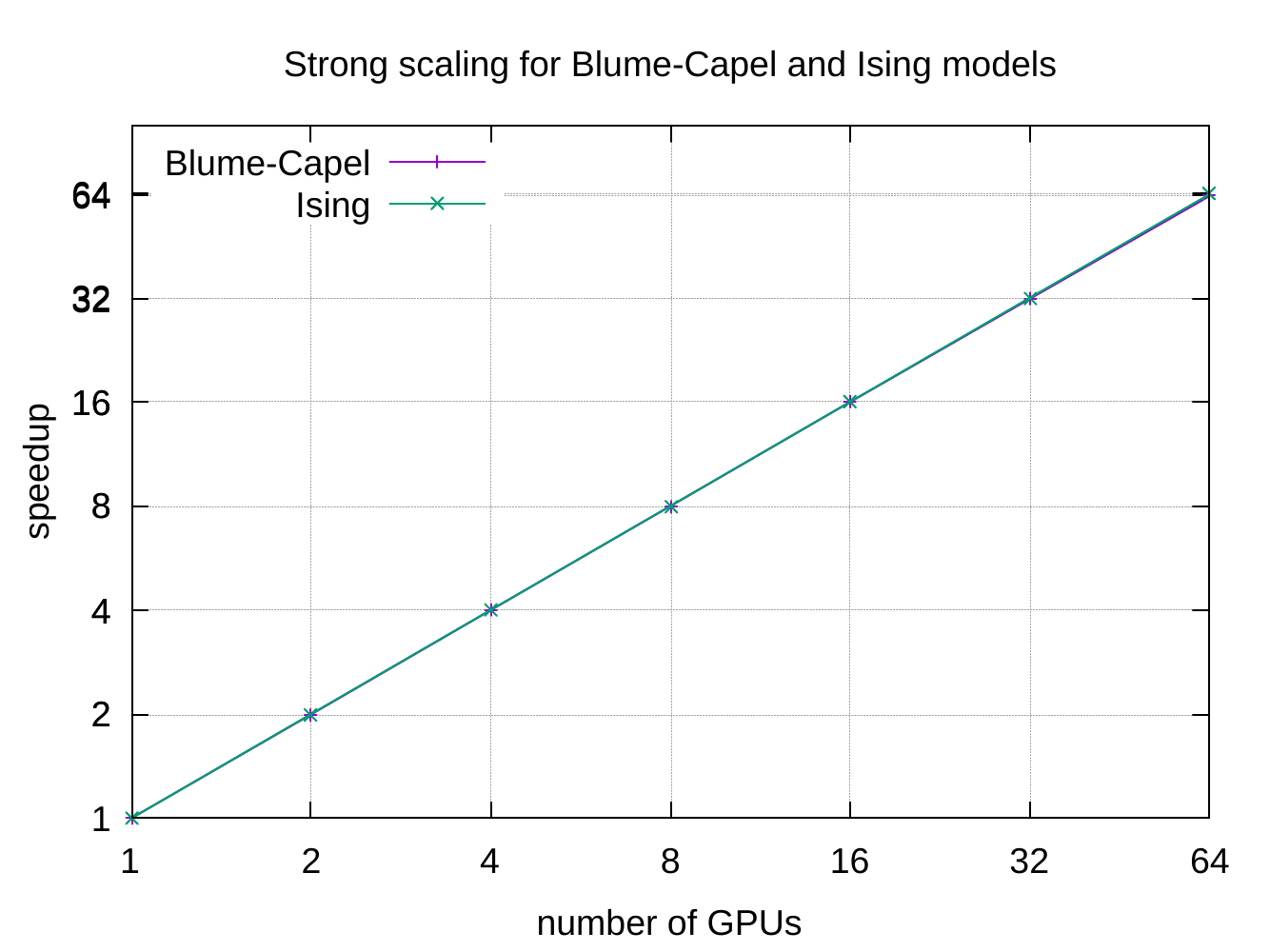}
    \end{minipage}
    \captionof{figure}{Plot showing the strong scaling performance of the Ising and Blume-Capel codes, based on the data from Table~\ref{tbl:strong_scaling_both}.}
    \label{fig:strong_scaling_both}
    \hfill\vline\hfill
    \begin{minipage}{\linewidth}
        \centering
        \begin{tabular}{ c r r r r }
            \toprule\midrule[0.1pt]
             \mr{2}{GPUs} & \mc{2}{c}{Blume-Capel ($L=2^{19}$)} & \mc{2}{c}{Ising ($L=2^{20}$)} \\ \cmidrule(lr){2-3}\cmidrule(lr){4-5}
                          &     updates/ns  &     runtime (sec) &   updates/ns  & runtime (sec) \\ \cmidrule(lr){1-1}\cmidrule(lr){2-3}\cmidrule(lr){4-5}
                        1 &             948 &             37.26 &          1802 &         78.87 \\
                        2 &            1890 &             18.69 &          3604 &         39.44 \\
                        4 &            3778 &              9.35 &          7196 &         19.75 \\
                        8 &            7547 &              4.68 &         14349 &          9.91 \\
                       16 &           15054 &              2.35 &         28678 &          4.96 \\
                       32 &           29910 &              1.18 &         57201 &          2.48 \\
                       64 &           59138 &              0.60 &        113813 &          1.25 \\
            \bottomrule
        \end{tabular}
    \end{minipage}
    \captionsetup{type=table}
    \caption{Strong scaling measurements of the Ising and Blume-Capel codes for the Monte Carlo kernel, using up to 64 GB200 GPUs with fixed-size systems, are presented. The Blume-Capel code was run with $L=2^{19}$ ($\sim2.8\times 10^{11}$ spins), while the Ising code with $L=2^{20}$ ($\sim 1.1\times 10^{12}$ spins), resulting in a total memory usage of $128$ GB in both cases. Each experiment was run for $128$ time steps. The runtime column includes both the Monte Carlo simulation time and the time spent on consistency checks, such as verifying that the total number of spins remains unchanged and computing the Schwinger-Dyson values.}
    \label{tbl:strong_scaling_both}
\end{table}

Table~\ref{tbl:weak_scaling_both} showcases the weak scaling measurements, where the system size per GPU is kept fixed while scaling from $1$ to $64$ GPUs. As in the strong scaling experiments, the lattice size per GPU was chosen to maximize memory usage. Specifically, for the Blume-Capel model, the system size per GPU was maintained at $2^{38}$ spins, while for the Ising model, it was set to $2^{40}$ spins. Figure \ref{fig:weak_scaling_both} shows the corresponding efficiency plot. As in the strong scaling case, the flip update rate was measured over 128 simulation steps. Although increasing the number of GPUs results in larger system boundaries (two rows of $L$ spins) while keeping the number of spins per device constant, the relative ratio of boundary to total spins remains small (with 64 GPUs, each device exchanges $2$ MB and $4$ MB of data with its two neighbors for the Ising and Blume-Capel models, respectively). Consequently, access to remote GPU memory has no significant impact on performance.

\begin{table}[h]
    \centering
    \begin{minipage}{\linewidth}
        \centering
        \includegraphics[scale=0.4]{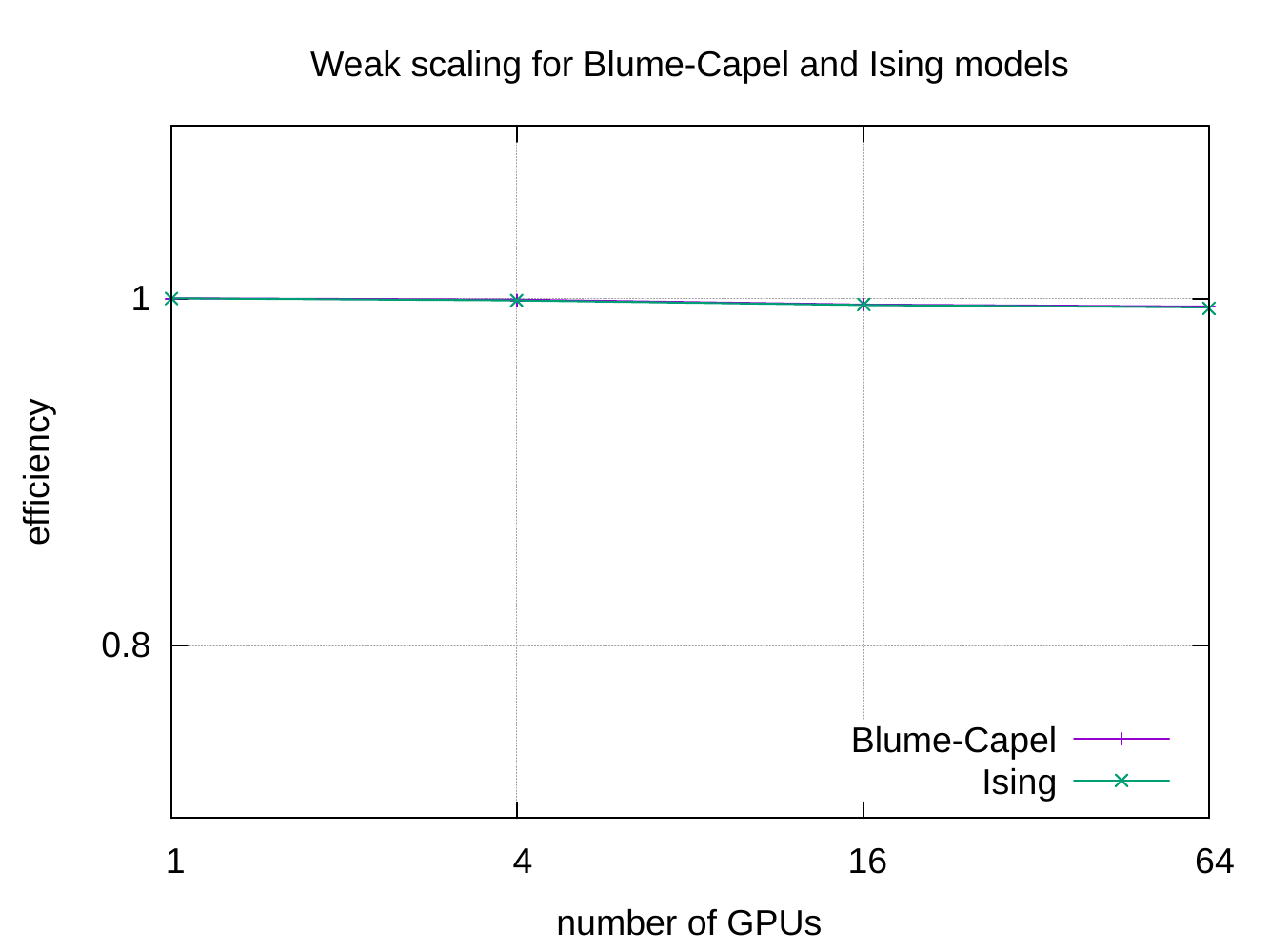}
    \end{minipage}
    \captionof{figure}{Plot showing the weak scaling performance of the Ising and Blume-Capel codes, based on the data from Table~\ref{tbl:weak_scaling_both}.}
    \label{fig:weak_scaling_both}
    \hfill\vline\hfill
    \begin{minipage}{\linewidth}
        \centering
        \begin{tabular}{ c c c r r c c r r}
            \toprule\midrule[0.1pt]
            \mr{2}{GPUs} &             \mc{4}{c}{Blume-Capel}                &               \mc{4}{c}{Ising}                    \\ 
            \cmidrule(lr){2-5}\cmidrule(lr){6-9}
                         &       $L$ & size (GB) &    upd/ns & runtime (sec) &       $L$ & size (GB) &     upd/ns & runtime (sec) \\ 
            \cmidrule(lr){1-1}\cmidrule(lr){2-5}\cmidrule(lr){6-9} 
                       1 & $2^{19}$  &       128 &       948 &         37.68 & $2^{20}$  &       128 &       1802 &         78.87 \\
                       4 & $2^{20}$  &       512 &      3789 &         37.76 & $2^{21}$  &       512 &       7199 &         78.97 \\
                      16 & $2^{21}$  &      2048 &     15113 &         37.82 & $2^{22}$  &      2048 &      28722 &         79.17 \\
                      64 & $2^{22}$  &      8192 &     60381 &         37.89 & $2^{23}$  &      8192 &     114729 &         79.29 \\
            \bottomrule
        \end{tabular}
    \end{minipage}
    \hfill
    \captionsetup{type=table}
    \caption{Weak scaling measurements of the Ising and Blume-Capel codes for the Monte Carlo kernel, using up to $64$ GB200 GPUs while keeping the system size per GPU fixed, are presented. The Blume-Capel code was run with $2^{38}$ spins per GPU, and the Ising code with $2^{40}$ spins per GPU, resulting in a total of
    $\sim1.8\times 10^{13}$ spins for the Blume-Capel model and 
    $\sim7.0\times 10^{13}$ spins for the Ising model (with $64$ GPUs). Each experiment was run for $128$ time steps. The runtime column includes both the Monte Carlo simulation time and the time spent on consistency checks, such as verifying that the total number of spins remains unchanged and computing the Schwinger-Dyson values.}
    \label{tbl:weak_scaling_both}
\end{table}

The highest spin update rates we measured with our codes are $114729$ and $60381$ updates/ns for the Ising and Blume-Capel implementations, respectively. To the best of our knowledge, the previous highest Ising flip rate was reported in Ref.~\cite{Cerebras}, where the authors achieved $61853$ updates/ns by running $754$ parallel simulations on a WSE, each with a size of $11586\times 16384$. Our results demonstrate that a GB200 NVL72 system---comparable to the WSE---delivers nearly twice the performance for Ising simulations ($\sim1.85\times$). Additionally, the GB200 NVL72 enables the study of significantly larger systems (up to $\sim370000\times$ larger), offering much higher statistical quality, since statistical errors scale inversely with system size in these simulations.  For example, a single run on a system of size $L=2^{22}$ is free of finite-size artifacts, provided the coherence length remains below $\xi(t)\approx 2^{18}$. In Ref.~\cite{PRLwannabe}, we simulated up to $\xi(t)\sim 2^{10}$, which would likely cause incipient finite-size artifacts in a system of linear dimensions $11586\times 16384$. Furthermore, the statistical errors for a single run of $L=2^{22}$ are roughly equivalent to those obtained from $2^{44}/(11586\times 16384)\approx 92675$ independent runs of a $11586\times 16384$ system.

It is worth noting that, with regard to the Blume-Capel model, we are unaware of any performance-focused implementations in the literature. To the best of our knowledge, this work presents the first high-performance implementation of the model.

\section{Conclusions \label{sec:conclusions}}

In this study, we presented high-performance implementations of the two-dimensional Ising and Blume-Capel models, optimized for large-scale multi-GPU simulations. These implementations leverage multi-node NVLink systems, such as the NVIDIA GB200 NVL72. By employing advanced memory management techniques and optimizing Monte Carlo kernels, our implementations achieve unprecedented simulation scales, handling up to $2^{46}$ spins for the Ising model and $2^{44}$ spins for the Blume-Capel model.
Our benchmarks demonstrate near-perfect strong and weak scaling up to 64 GPUs, highlighting the effectiveness of the NVLink interconnect in enabling memory sharing across multiple compute nodes. The Ising model achieves a peak update rate of $114729$ updates/ns, significantly surpassing previous performance records. Meanwhile, the Blume-Capel model reaches $60381$ updates/ns, marking the first high-performance implementation of this model to our knowledge. These results further demonstrate the versatility of GPUs as general-purpose accelerators, showing their ability to scale seamlessly to rack-level configurations. Advances in high-speed interconnects, unified memory architectures, and parallel programming frameworks now allow programmers to treat distributed GPUs as a unified compute resource.

A natural direction for future developments would be to extend the codes to support 3D systems. Although the shared buffer based on fabric memory and the optimized random bit generation could be reused with minimal modifications, the spin processing kernels would likely require substantial changes. A 3D lattice may benefit from a different spin layout in memory, which would necessitate a new memory access pattern for reading neighboring spins. Furthermore, careful consideration should be given to the domain decomposition strategy to minimize intra-GPU communications when computing correlations.

Beyond raw performance, our implementations enable simulations of significantly larger systems than previously possible, improving the statistical reliability of results and minimizing finite-size effects. The custom correlation function protocol and Schwinger-Dyson checks further ensure the correctness and consistency of our simulations. These advances open up new opportunities for large-scale statistical physics studies, including investigations of critical dynamics, universality classes, and non-equilibrium phenomena at previously unattainable scales.

\section*{Acknowledgements}

This work was partially supported by MCIN/AEI/10.13039/501100011033 and by
``ERDF A way of making Europe'' through Grant No. PID2022-136374NB-C21.
The work of Nikolaos G. Fytas and Alexandros Vasilopoulos was supported by the Engineering and Physical Sciences Research Council (grant EP/X026116/1 is acknowledged). 
Isidoro Gonz\'{a}lez-Adalid Pemart\'{i}n and Massimo Bernaschi acknowledge the support of the National Center for HPC, Big Data and Quantum Computing, Project CN\_00000013 – CUP E83C22003230001 and CUP B93C22000620006, Mission 4 Component 2 Investment 1.4, funded by the European Union – NextGenerationEU. 
\bibliographystyle{unsrt}  

\bibliography{CPCbiblio}

\end{document}